\documentclass[a4paper,fleqn]{revtex4}
\begin{document}

\title{Adiabatic continuity and broken symmetry in many-electron systems:\\ a variational perspective}

\author{D.\ Baeriswyl}
\affiliation{Department of Physics, University of Fribourg, 
CH-1700 Fribourg, Switzerland}

\vskip 0.5cm
\centerline{\it Dedicated to Dieter Vollhardt on the occasion of his 60th birthday.}
\vskip 0.5cm
\begin{abstract}
Variational wave functions are very useful for describing the panoply of ground states found in interacting many-electron systems. Some particular trial states are ``adiabatically'' linked to a reference state, from which they borrow the essential properties. A prominent example is the Gutzwiller ansatz, where one starts with the filled Fermi sea. A simple soluble example, the anisotropic XY chain, illustrates the adiabatic continuity of this class of wave functions. To describe symmetry breaking, one has to modify the reference state accordingly. 
Alternatively, a quantum phase transition can be described by a pair of variational wave functions, starting at weak and strong coupling, respectively. 
 \end{abstract}
\maketitle                
\section{Introduction}
According to the adiabatic theorem of quantum mechanics a system governed by a Hamiltonian with time-dependent parameters evolves from an initial eigenstate to an eigenstate at later times if the parameters vary infinitesimally slowly with time \cite{Born28, Kato50}. An adiabatic evolution from an initial non-interacting eigenstate to an interacting state by slowly switching on the interactions is also a key element in Landau's theory of the Fermi liquid \cite{Landau57, Pines66}. 
Anderson generalized this idea to the concept of adiabatic continuity, which basically means the continuous mapping of a simple reference state -- the non-interacting Fermi sea in Landau's theory -- to a more complex state which however shares the essential properties with the parent state \cite{Anderson84}. This can only work for systems which do not experience any symmetry breaking on the way. It is the aim of this paper to show that a similar continuity characterizes a whole class of variational wave functions. When these are applied without precaution to systems that undergo some phase transition one may easily be led to wrong conclusions. 

Section 2 recalls the well-known example of the Gutzwiller wave function, which represents a metallic state, even for a half-filled band. It can therefore be viewed as a  microscopic realization of Landau's phenomenological theory of the Fermi liquid. Section 3 presents the exactly solvable anisotropic XY chain which exhibits a quantum phase transition for an infinitesimal anisotropy. A Gutzwiller-type variational ansatz fails in signaling this transition. Section 4 deals with various kinds of quantum phase transitions, both with and without symmetry breaking. ``Dual'' pairs of variational wave functions give an appealing picture of metal-insulator transitions. In Section 5 a different class of wave functions is briefly mentioned, which is able to promote long-range order without forcing symmetry breaking in the reference state. 
\section{Adiabatic continuity of the Gutzwiller ansatz}
When Martin Gutzwiller introduced his famous ansatz for the ground state of a correlated electron system he wanted ``to present a new approach to the problem of ferromagnetism in a metal''
\cite{Gutz63}. As a particular example he considered the Hubbard Hamiltonian
\begin{equation}
H=-t\sum_{\langle n,m\rangle, \sigma}
(c_{n\sigma}^\dag c_{m\sigma}+c_{m\sigma}^\dag c_{n\sigma})
+U\sum_n c_{n\uparrow}^\dag c_{n\uparrow}c_{n\downarrow}^\dag c_{n\downarrow}\, ,
\end{equation}
where $c_{n\sigma}^\dag$ ($c_{\sigma}$) creates (annihilates) an electron with spin $\sigma$ at site $n$ . The issue was not to find a good trial state for ferromagnetism, because the lowest-energy fully polarized ferromagnetic state for $N$ electrons is simply constructed by occupying the $N$ lowest levels with only up (or down) electrons, but rather to show that no other state has lower energy. This problem has been intensively studied ever since Gutzwiller's work and is still not completely understood \cite{Fazekas99}. Only for special lattices and/or band structures it can be rigorously shown that the ground state of the Hubbard model is ferromagnetic 
\cite{Gulacsi08}.

Gutzwiller proposed the ansatz
\begin{equation}
\vert\Psi(g)\rangle=\exp\left[-g\sum_n
c_{n\uparrow}^\dag c_{n\uparrow}c_{n\downarrow}^\dag c_{n\downarrow}\right]\, \vert\Psi_0\rangle
\label{Gutzwiller}
\end{equation}
for the correlated non-magnetic state, where $\vert\Psi_0\rangle$ is the filled Fermi sea. The variational parameter $g$ reduces double occupancy and thus the interaction energy. As to the kinetic energy, a simple approximate expression can be obtained by neglecting the dependence of the hopping processes on the detailed spin configurations \cite{Gutz63, Ogawa75}. This ``Gutzwiller approximation'' allows the calculation of physical quantities such as the momentum distribution and, by including a Zeeman term,  the magnetic susceptibility. For a generic band filling, {\it i.e.}, away from half filling, correlation effects essentially increase the effective mass but do not destroy the metallicity. An interesting link to Landau's Fermi-liquid theory has been established by Vollhardt \cite{Vollhardt84}, who was able to establish a direct connection between microscopic quantities and Landau parameters. The smooth connection between the correlated state $\vert\Psi(g)\rangle$ and the uncorrelated state $\vert\Psi_0\rangle$ in Eq. (\ref{Gutzwiller}) has thus its counterpart in the adiabatic hypothesis underlying Landau's Fermi-liquid theory. 

The special density of one electron per site (half-filled band) is worth being discussed in some detail, because in this case the Gutzwiller approximation predicts a metal-insulator transition at a critical value $U_c=8\vert\varepsilon_0\vert$, where $\varepsilon_0$ is the kinetic energy per site of the uncorrelated system  \cite{Brinkman70}. Signatures of the transition are both the vanishing of double occupancy and the disappearance of the Fermi step  in the momentum distribution function. However, more accurate treatments of the Gutzwiller ansatz obtain this transition only for 
$U\rightarrow\infty$ and thus show that the variational state (\ref{Gutzwiller}) remains metallic for all finite values of $U$. This is clearly seen in the case of the one-dimensional Hubbard model for which the Gutzwiller ansatz can be handled exactly \cite{Metzner87, Gebhard87}, but there exists also a simple argument why this remains true for any (finite) dimension. The argument is based on the Drude weight which is related to the sensitivity of the ground state with respect to changes in the boundary conditions \cite{Kohn64}. The Drude weight is finite for the Gutzwiller ground state for any finite value of $U$, even at half filling \cite{Millis91, Dzierzawa97}.
\section{Quantum phase transition for infinitesimal coupling: The anisotropic XY chain}
A simple example illustrates the smooth link between a variational wave function of the Gutzwiller type and its reference state, the ``bare'' ground state. We consider the anisotropic XY chain, represented by the Hamiltonian
\begin{equation}
H=H_0+\gamma H'\, ,
\end{equation}
where
\begin{eqnarray}
H_0&=&\sum_n\left(S_n^{(x)}S_{n+1}^{(x)}
+S_n^{(y)}S_{n+1}^{(y)}\right)\, ,\nonumber\\
H'&=&\sum_n\left(S_n^{(x)}S_{n+1}^{(x)}
-S_n^{(y)}S_{n+1}^{(y)}\right)\, .
\end{eqnarray}
The parameter $\gamma$, $0\le\gamma\le 1$, interpolates between the isotropic XY chain 
($\gamma=0$) and the one-dimensional Ising model ($\gamma=1$). This Hamiltonian has been
diagonalized exactly by Lieb, Schultz and Mattis \cite{Lieb61}. The main steps of the solution are reproduced in the appendix. It turns out that this model exhibits a quantum phase transition at
$\gamma =0$, from a ground state without long-range order at $\gamma =0$ to a state with
long-range antiferromagnetic order of the $x$-components of the spins for
any $\gamma >0$ \cite{Lieb61}. The transition reveals itself also in the singularity of the ground-state energy (per site, $\gamma\ll 1$)
\begin{equation}
\varepsilon_0\, \sim\, 
-\frac{1}{\pi}\left[1+\frac{\gamma^2}{2}\left(\log{\frac{4}{\gamma}}-\frac{1}{2}\right)\right]\, .
\label{exact}
\end{equation}
Another interesting quantity is the overlap $\langle\Psi(\gamma') \vert\Psi(\gamma)\rangle$  between ground states of two Hamiltonians with different anisotropies \cite{Zanardi06}. We consider in particular the overlap at criticality, i.e. for $\gamma'=0$ and $\gamma\ll 1$. It is convenient to introduce the fidelity susceptibility per site \cite{You07}
\begin{equation}
\chi_F=
\lim_{\gamma\rightarrow 0}\frac{-2}{L\gamma^2}\log \langle\Psi(0)\vert\Psi(\gamma)\rangle\, .
\end{equation}
As shown in the appendix, we find for the XY chain for $\gamma\ll 1$
\begin{equation}
-\frac{2}{L\gamma^2}\log \langle\Psi(0)\vert\Psi(\gamma)\rangle
=\frac{\pi -2}{2\pi\gamma}-\frac{1}{8}\, ,
\end{equation}
which shows that $\chi_F$ diverges, as expected for a quantum-critical point  \cite{Gu10}.

We turn now to a variational ansatz \`a la Gutzwiller. The transformation to fermion operators
$c_n,\, c_n^\dag$ (appendix) leads to the expressions
\begin{eqnarray}
H_0&=&\frac{1}{2}\sum_n(c_n^\dag c_{n+1}+c_{n+1}^\dag c_n)\, ,\nonumber\\
H'&=&\frac{1}{2}\sum_n(c_n^\dag c_{n+1}^\dag+c_{n+1}c_n)\, ,
\end{eqnarray}
or, after Fourier transformation,
\begin{eqnarray}
H_0&=&\sum_k\cos k\, c_k^\dag c_k\, ,\nonumber\\
H'&=&i\sum_{k>0}\sin k\, (c_k^\dag c_{-k}^\dag-c_{-k}c_k)\, .
\end{eqnarray}
The Gutzwiller ansatz
\begin{equation}
\vert\Psi(g)\rangle = e^{-gH'}\vert\Psi_0\rangle\, ,
\end{equation}
where $\vert\Psi_0\rangle$ is the ground state of $H_0$, {\it i.e.}, all levels with $|k|>\pi/2$ are
occupied and all others are empty, can be written 
\begin{equation} \label{trial}
\vert\Psi(g)\rangle =\prod_{0\le k<\pi/2}
(\cosh \vartheta_k-i\sinh\vartheta_kc_k^\dag c_{-k}^\dag) \prod_{k>\pi/2}
(\cosh \vartheta_k+i\sinh\vartheta_kc_{-k}c_k)\, \vert\Psi_0\rangle\, ,
\end{equation}
where $\vartheta_k=g\sin k$. The expectation value of the Hamiltonian is readily calculated,
\begin{equation}
\frac{\langle\Psi(g)\vert H\vert\Psi(g)\rangle}{\langle\Psi(g)\vert\Psi(g)\rangle}
=-\sum_{k\ge 0}\left\{\frac{|\cos k|}{\cosh(2g\sin k)}+\gamma\sin k\tanh(2g\sin k)\right\}\, .
\end{equation}
The minimization of this expression for $g\ll 1$ gives $g=(3\pi/8)\gamma$ and
an energy per site
\begin{equation}
\varepsilon_{\mbox{\tiny min}}\, \sim\, 
-\frac{1}{\pi}\left(1+\frac{3\pi^2}{32}\gamma^2\right)\, .
\end{equation}
This is an analytic function of $\gamma$, in contrast to the exact result, Eq. (\ref{exact}).
A striking difference is also obtained for the overlap. The (normalized) trial state of Eq. (\ref{trial}) leads to
\begin{equation}
\frac{\langle\Psi(0)\vert\Psi(g)\rangle}{\sqrt{\langle\Psi(g)\vert\Psi(g)\rangle}}=
\prod_{k>0}\frac{\cosh (g\sin k)}{\sqrt{\cosh (2g\sin k)}}
\,\sim\, \prod_{k>0}(1-\frac{1}{2}g^2\sin^2k)\, 
\, \sim\, e^{-\frac{Lg^2}{8}}\, .
\end{equation}
Thus in the limit $\gamma\rightarrow 0$ the fidelity susceptibility tends to a constant for the Gutzwiller ansatz,
\begin{equation}
\chi_F\rightarrow \left(\frac{3\pi}{16}\right)^2\approx 0.35,
\end{equation} 
while it diverges $\sim \gamma^{-1}$ for the exact ground state.
\section{Variational wave functions and quantum phase transitions}
The discrepancy found for the anisotropic XY chain between the exact and variational ground states exists also in the case of the one-dimensional Hubbard model. At half filling the exact
ground state energy \cite{Lieb68} is not analytic for $U\rightarrow 0$ \cite{Takahashi71, Metzner89}, while the Gutzwiller ansatz yields an analytic behavior \cite{Horsch81}. It is natural to attribute this difference to the fact that a metal-insulator transition occurs in the exact ground state at $U=0$, but not so in the Gutzwiller ansatz, as explained above.  We can remedy this problem by modifying the reference state $\vert\Psi_0\rangle$, which so far was considered to be metallic. We could introduce an alternating spin density, as predicted by the Unrestricted Hartree-Fock approximation. A more interesting possibility is bond alternation in the sense of an alternating bond order. This instability does not occur in a simple mean-field treatment, which would require a finite electron-phonon coupling (Peierls instability). However, if the reference state $\vert\Psi_0\rangle$
is taken as a mean-field state with alternating bond order, the (Gutzwiller) variational energy is lowered, even in the absence of electron-phonon coupling. The condensation energy due to this symmetry breaking is given by the non-analytic term \cite{Baeriswyl85}
\begin{equation}
\varepsilon_{cond}\sim -t\, e^{-3.85(4t/U)^2}\, .
\end{equation}
Long-range order is of course not expected to survive if quantum fluctuations are fully taken into account, rather one expects the correlation functions, both for spin and for the bond order, to
decay algebraically with distance.

In two dimensions fluctuations are less severe and therefore long-range antiferromagnetic order, as found consistently for the two-dimensional Hubbard model at half filling by using simple mean-field theory or more elaborate methods, is commonly believed to resist quantum fluctuations. A more subtle question is the issue of superconductivity because, similarly to bond alternation in one dimension, a superconducting instability does not occur within simple mean-field theory. For certain electron densities the Gutzwiller ansatz, linked to a reference state with $d$-wave pairing, is found to have lower energy than the state without symmetry breaking (or with an antiferromagnetic reference state) \cite{Giamarchi91}. This remains true for more sophisticated variational ground states \cite{Yokoyama04, Eichenberger07, Baeriswyl09}. Whether the exact ground state exhibits superconductivity remains an open issue \cite{Aimi07}, but the variational results indicate at least strong superconducting correlations.

We turn now to the Mott phenomenon, which {\it a priori} does not involve any symmetry breaking,
but rather is a topological (quantum) phase transition as a function of coupling strength, from a metallic to an insulating state. The transition typically occurs in a region where the interaction strength is of the order of the band width. In contrast to the instabilities discussed above, the Mott metal-insulator transition cannot be described in the framework of the Gutzwiller ansatz alone, which is metallic in the absence of symmetry breaking, but rather calls for a complementary variational state for the insulating side of the transition. In the case of the Hubbard model the following ansatz has been proposed for the ground state of the insulating phase \cite{Baeriswyl87},
\begin{equation}
\vert\Psi(h)\rangle=\exp\left[-h\sum_{\langle n,m\rangle, \sigma}
(c_{n\sigma}^\dag c_{m\sigma}+c_{m\sigma}^\dag c_{n\sigma})\right]\vert\Psi_\infty\rangle\, ,
\label{DB}
\end{equation} 
where $\vert\Psi_\infty\rangle$ is the exact ground state for $U\rightarrow\infty$. This state, the dual partner of the Gutzwiller ansatz \cite{Dzierzawa95}, is also adiabatically connected to its reference state $\vert\Psi_\infty\rangle$. The Drude weight vanishes at half filling \cite{Dzierzawa97} and therefore 
$\vert\Psi(h)\rangle$ does represent an insulating state. Unfortunately, it is very hard to handle $\vert\Psi_\infty\rangle$, which is the ground state of the Heisenberg antiferromagnet at half filling and of the $t$-$J$ Hamiltonian close to half filling, except for some special cases, such as the $1/r$ Hubbard chain \cite{Dzierzawa95} or for infinite dimensions \cite{Dzierzawa97}. In these cases
a variational procedure using two trial states, one linked to the ground state at $U=0$, the other to the ground state for $U\rightarrow\infty$, is rather successful in predicting the location of the Mott transition \cite{Dzierzawa97}. On the other hand, the nature of the transition is inevitably found to be of first order, also in cases where it is known to be continuous  \cite{Dzierzawa95}. 


A related phenomenon is the crystallization of electrons, proposed by Wigner for the dilute three-dimensional electron gas \cite{Wigner38} and later by Hubbard for narrow-band solids
\cite{Hubbard78}. In this case the metal-insulator transition leading to the Wigner crystal involves a reduction of translational symmetry. For the case of electrons on a lattice the trial state of Eq. (\ref{DB}) can again be used, this time for describing the crystalline phase, starting from the classical configuration which minimizes the Coulomb energy \cite{Valenzuela03}. This leads to a quantum crystal where electrons are preferentially located at the sites of the reference configuration, but occasionally visit neighboring sites. With increasing relative strength of the kinetic energy these excursions become more and more frequent until the crystal melts. It would be interesting to test this two-sided variational procedure for the anisotropic XY chain by studying the dual partner of the Gutzwiller ansatz, starting from the N\'eel state, the ground state of the antiferromagnetic Ising model. 
\section{Discussion}
The variational ground states presented above are all smoothly linked to some reference state. The case of a transition occurring at a large value of the coupling constant, such as the Mott transition and the Wigner crystallization, is rather well described by a pair of wave functions that are linked to the ground states for vanishing and infinite coupling strengths, respectively. In this way the critical coupling strength may be determined rather accurately, but the nature of the transition is not described reliably. Therefore more refined methods are required for treating the region of the transition, which may be of second order, in contrast to the prediction in terms of two dual wave functions, or it may even be split into two critical points with an intermediate phase in-between, as debated in the case of Wigner crystallization \cite{Kivelson05, Ceperley09}.

A single variational state of the type discussed above does not change the qualitative nature of a
reference state. For instance, these states do not lead to long-range order not included already in the reference state $\vert\Psi_0\rangle$, nor do they destroy long-range order if it is present in
$\vert\Psi_0\rangle$. To achieve such a radical change due to a continuous change of some variational parameter, a different class of variational states has to be used, such as the 
resonance-valence-bond state with a variable bond-length distribution \cite{Liang88}. This state is able to describe both short-range and long-range antiferromagnetic order on a square lattice. 
It is also possible to describe the Mott transition using a single variational wave function, by adding a Jastrow factor correlating distant electron densities \cite{Capello05}.
\acknowledgments{I have profited from discussions with M. Menteshashvili.}

\def\bstname{adp}
\begin{appendix}
\section{Anisotropic XY model in one dimension}
We consider the anisotropic XY chain defined by the Hamiltonian
\begin{equation}
H=\sum_{n=1}^L\left[(1+\gamma)S_n^{(x)}S_{n+1}^{(x)}
+(1-\gamma)S_n^{(y)}S_{n+1}^{(y)}\right]\, ,
\end{equation}
where $S_n^{(x)},\, S_n^{(y)}$ are spin $1/2$ operators and $0\le\gamma\le 1$. We assume
periodic boundary conditions, $S_{L+1}^{(\alpha)}=S_1^{(\alpha)}$, $\alpha=x,y$, and choose
$L=4M+2$ (where $M$ is a positive integer). The Jordan-Wigner transformation
\begin{equation}
S_n^{(x)}+iS_n^{(y)}=c_n^\dag e^{i\pi\sum_{j=1}^{n-1}c_j^\dag c_j}\, ,
\end{equation}
where $c_n^\dag,\, c_n$ are fermionic creation and annihilation operators, yields a simple
quadratic expression for the Hamiltonian,
\begin{equation}
H=\frac{1}{2}\sum_{n=1}^L(c_n^\dag c_{n+1}+\gamma c_n^\dag c_{n+1}^\dag +\mbox{h.c.})\, .
\end{equation}
Introducing the Fourier transform
\begin{equation}
c_n=\frac{1}{\sqrt{L}}\sum_k e^{ikn}c_k, \, k=\frac{2\pi}{L}\nu,\, -2M\le\nu\le 2M+1,
\end{equation}
we rewrite the Hamiltonian
\begin{equation}
H=\sum_k \cos{k}\, c_k^\dag c_k + i\gamma\sum_{k>0}\sin{k}\, (c_k^\dag c_{-k}^\dag -c_{-k} c_k)
\end{equation}
and diagonalize it using the Bogoliubov transformation
\begin{eqnarray}
c_k&=&e^{i\pi/4}(\cos{\varphi_k}\, d_k +\sin{\varphi_k}\, d_{-k}^\dag)\, ,\nonumber\\
c_{-k}^\dag &=&e^{-i\pi/4}(-\sin{\varphi_k}\, d_k +\cos{\varphi_k}\, d_{-k}^\dag)\, ,
\end{eqnarray}
where 
\begin{equation}
\tan(2\varphi_k)=-\gamma\tan{k}\, .
\end{equation}
This yields the excitation spectrum 
\begin{equation}
E_k=\sqrt{1-(1-\gamma^2)\sin^2{k}}\, .
\end{equation}
With the choice
\begin{equation}
\cos\varphi_k=\left(\frac{E_k+\cos k}{2E_k}\right)^\frac{1}{2}\, ,\qquad 
\sin\varphi_k=-\left(\frac{E_k-\cos k}{2E_k}\right)^\frac{1}{2}
\label{choice}
\end{equation}
the Hamiltonian reads
\begin{equation}
H=\sum_k E_k\left(d_k^\dag d_k-\frac{1}{2}\right)\, .
\end{equation}
The ground state energy per site for $L\rightarrow\infty$ is given by
\begin{equation}
\varepsilon_0=-\frac{1}{2L}\sum_k E_k\rightarrow -\frac{1}{\pi}E(1-\gamma^2)\, \sim\, 
-\frac{1}{\pi}-\frac{\gamma^2}{2\pi}\left(\log{\frac{4}{\gamma}}-\frac{1}{2}\right)\, ,
\end{equation}
where $E(1-\gamma^2)$ is the complete elliptic integral. 
The ground state $\vert\Psi(\gamma)\rangle$, defined by $d_k\vert\Psi(\gamma)\rangle =0\, ,$
can be written
\begin{equation}
\vert\Psi(\gamma)\rangle=
\prod_{k>0}(\cos\varphi_k+i\sin\varphi_k\, c_k^\dag c_{-k}^\dag)\vert0\rangle\, ,
\end{equation}
where $\vert 0\rangle$ is the vacuum of $c$-particles. The overlap between two ground states for different anisotropies $\gamma, \gamma'$ is given by
\begin{equation}
\langle\Psi(\gamma')\vert\Psi(\gamma)\rangle=\prod_{k>0}\cos (\varphi_k-\varphi'_k)\, .
\end{equation}
We are particularly interested in the overlap between the ground state of the isotropic XY model 
($\gamma=0$) and that of the anisotropic model ($\gamma>0$). With the choice (\ref{choice}) we get
\begin{equation}
\langle\Psi(0)\vert\Psi(\gamma)\rangle=\prod_{k>0}\left(\frac{E_k+|\cos k|}{2E_k}\right)^\frac{1}{2}\, ,
\end{equation}
i.e., in the thermodynamic limit,
\begin{equation}
\log \langle\Psi(0)\vert\Psi(\gamma)\rangle
=\frac{L}{2\pi}\int_0^{\frac{\pi}{2}}dk\, \log\frac{E_k+\cos k}{2E_k}\, . 
\end{equation}
To obtain the asymptotic behavior for $\gamma\rightarrow 0$ we have to distinguish between the region of $k$ values close to $\frac{\pi}{2}$ and the other region in $k$ space. In the latter we can readily expand the integrand in powers of $\gamma$, while in the former region we replace $\cos k$ by $\frac{\pi}{2}-k$ and then integrate that part exactly. Summing up the two contributions we
find
\begin{equation}
\log \langle\Psi(0)\vert\Psi(\gamma)\rangle 
\, \sim\, \frac{L}{2\pi}\left[-\left(\frac{\pi}{2}-1\right)\gamma+\frac{\pi}{8}\gamma^2\right]\, .
\end{equation}
\end{appendix}
%
%

\end{document}